\documentclass{article} 
\usepackage{iclr2019_conference,times}

\usepackage{amsmath,amsfonts,bm}

\def\eqref#1{equation~\ref{#1}}

\def\1{\bm{1}}

\DeclareMathAlphabet{\mathsfit}{\encodingdefault}{\sfdefault}{m}{sl}
\SetMathAlphabet{\mathsfit}{bold}{\encodingdefault}{\sfdefault}{bx}{n}

\usepackage{hyperref}
\usepackage{url}
\usepackage{amsmath}
\usepackage{amsfonts}
\usepackage{amssymb}
\usepackage{booktabs}
\usepackage{multirow}
\usepackage{url}

\newcommand{\norm}[1]{\left\lVert#1\right\rVert}

\title{A note on hyperparameters in black-box adversarial examples}

\author{Jamie Hayes \\
University College London\\
\texttt{j.hayes@cs.ucl.ac.uk} 
}

\iclrfinalcopy 
\begin{document}

\maketitle

\begin{abstract}

Since \cite{biggio2013evasion} and \cite{szegedy2013intriguing} first drew
attention to adversarial examples, there has been a flood 
of research into defending and attacking machine learning models. 
However,
almost all proposed attacks assume white-box access
to a model. In other words, the attacker
is assumed to have perfect knowledge of
the model’s weights and architecture. With this insider
knowledge, a white-box attack can leverage gradient information 
to craft adversarial examples. Black-box attacks assume no knowledge of the model weights
or architecture. These attacks craft adversarial examples using information only contained in the logits or hard classification label. Here,
we assume the attacker can use the logits in order to find an adversarial example. Empirically, we show that 2-sided stochastic gradient estimation techniques 
are not sensitive to scaling parameters, and can be used to mount powerful black-box attacks requiring relatively few model queries.

\end{abstract}

\section{Gradient estimation methods}

Black-box attacks usually rely on either the \textit{transferability} property of adversarial
examples (\cite{papernot2016transferability}) or gradient estimation techniques. Gradient estimation
techniques such as finite differences
come at the expense of the number of model queries. To estimate the gradient
of a $d$ dimensional vector requires $2d$ model queries, since the gradient of each
dimension is measured independently. For this reason, recent work has explored
stochastic methods for approximating the true gradient (\cite{ilyas2018black, uesato2018adversarial}). Given a model $f$~\footnote{
For notational convenience, we represent both the model evaluation and loss function evaluation of an input with respect to a target label
by $f$.} and an input $x\in \mathbb{R}^{d}$, we estimate the gradient by one of the following methods:

\begin{gather}
\frac{1}{n}\sum\limits_{i=i}^n\bigg[\frac{f(x + \delta \Delta_i) - f(x)}{\delta}\bigg]\cdot \xi_i \label{eq: 1side}\\
\frac{1}{n}\sum\limits_{i=i}^n\bigg[\frac{f(x + \delta \Delta_i) - f(x - \delta \Delta_i)}{2\delta}\bigg]\cdot \xi_i \label{eq: 2side}
\end{gather}

\begin{table}[ht]
\caption{Attack results for different estimation methods.}
\begin{center}
\begin{tabular}{lccc}
\toprule 
\textbf{Attack} & \textbf{$\delta$} & \textbf{Success Rate (\%)} & \textbf{Median Queries} \\
\midrule 
\multirow{ 3}{*}{\textit{NES (2-sided)}}    & 1e-2  & 100               & 14815          \\
& 1e-3  & 100               & 17723          \\
& 1e-4  & 100               & 20222          \\
\cmidrule{2-4}
\multirow{ 3}{*}{\textit{RDSA (2-sided)}}  & 1e-2  & 100               & 13719          \\
& 1e-3  & 100               & 18564          \\
& 1e-4  & 100               & 19635        \\ 
\cmidrule{2-4}
\multirow{ 3}{*}{\textit{SPSA (2-sided)}}   & 1e-2  & 100               & 14586          \\
& 1e-3  & 100               & 17442          \\
& 1e-4  & 100               & 19916          \\
\cmidrule{2-4}
\multirow{ 3}{*}{\textit{SPSA (1-sided)}} & 1e-2  & 9.88              & 50206          \\
& 1e-3  & 91.59             & 23816          \\
& 1e-4  & 99.89             & \textbf{10244}         \\ 
\bottomrule 
\end{tabular}
\end{center}
\label{tab: results}
\end{table}


where $\delta \in \mathbb{R}$ is a small constant and $\xi_i, \Delta_i \in \mathbb{R}^d$ are random vectors sampled
from some distribution $P$. We observed that the choice of $P$ is largely unimportant; black-box
attack success for $P \sim \mathcal{N}(0, I)$ (\textsc{NES} (\cite{ilyas2018black})), $P \sim \text{Bernoulli}_{\pm 1}$ (\textsc{SPSA} (\cite{uesato2018adversarial})),
$P \sim \mathcal{U}(-1, 1)$ (\textsc{RDSA}) is approximately equivalent. Gradient estimation given by (\ref{eq: 1side}) is referred to as 1-sided, and
gradient estimation given by (\ref{eq: 2side}) is referred to as 2-sided. For all attacks, we take $\xi_i = \Delta_i^{-1}$, and the estimation can be viewed as finite differences on a 
random basis.

\section{Experiments}

We run the PGD attack (\cite{madry2017towards}) with gradient estimation, for $\epsilon = 0.05$ on the NIPS 2017 adversarial vision competition dataset (\cite{kurakin2018adversarial}).
This consists of 1000 Imagenet-like images of size $299 \times 299 \times 3$. For each image, we select the least likely class as the
adversarial example target class. We consider the attack successful if the predicted class is the target class and $\norm{x-x_{\textit{adv}}}_{\infty} < \epsilon$,
and the attack requires fewer than one million queries to the model. Table \ref{tab: results} shows the results for different choices of random directions.
For 2-sided attacks, all achieve perfect success rates in crafting adversarial examples, while exhibiting little sensitivity to the choice of $\delta$. For 1-sided \textsc{SPSA}, the attack is 
extremely sensitive to the choice of $\delta$; for a $\delta$ of $0.01$, the attack is successful fewer than one times in ten, while a $\delta$ of $0.0001$ the attack
has near perfect success rate and also requires on average only $10244$ queries, $3475$ fewer than the best 2-sided attack.

In conclusion, we found that the choice of random direction is largely unimportant in practical attacks. An attacker choosing a 1-sided perturbation may require fewer queries
to the model, however this is heavily dependent on the choice of $\delta$. Reproducible code can be found at {\small \texttt{\url{https://github.com/jhayes14/black-box-attacks}}}. 

\bibliography{iclr2019_conference}
\bibliographystyle{iclr2019_conference}

\end{document}